\documentclass[fleqn,10pt]{wlscirep}
\usepackage[T1]{fontenc}
\usepackage[utf8]{inputenc}
\usepackage{amsmath}
\usepackage{amssymb}
\usepackage{todonotes}
\usepackage{comment}

\usepackage{hyperref}
\hypersetup{
  colorlinks=true,
  allcolors=blue,
}

\title{Laser-driven vacuum breakdown waves}

\author[1,*]{A.~S.~Samsonov}
\author[1]{E.~N.~Nerush}
\author[1]{I.~Yu.~Kostyukov}
\affil[1]{Institute of Applied Physics of the Russian Academy of Sciences,
46 Ulyanov St., Nizhny Novgorod 603950, Russia}
\affil[*]{blendersamsonov@yandex.ru}

\begin{abstract}
It is demonstrated by three-dimensional quantum electrodynamics~---~particle-in-cell (QED-PIC) simulations that vacuum breakdown wave in the form of QED cascade front can propagate in an extremely intense plane electromagnetic wave. The result disproves the statement that the self-sustained cascading is not possible in a plane wave configuration. In the simulations the cascade initiates during laser-foil interaction in the light sail regime. As a result, a constantly growing electron-positron plasma cushion is formed between the foil and laser radiation. The cushion plasma efficiently absorbs the laser energy and decouples the radiation from the moving foil thereby interrupting the ion acceleration. The models describing propagation of the cascade front and electrodynamics of the cushion plasma are presented and their predictions are in a qualitative agreement with the results of numerical simulations.
\end{abstract}

\begin{document}

\flushbottom
\maketitle
\thispagestyle{empty}

\section*{Introduction}

With a fast progress in PW laser technology avalanche-like production of electron-positron pair plasma via QED cascades started to attract a great attention~\cite{Bell2008,fedotov2010limitations,elkina2011qed,Nerush2011,grismayer2016laser}. In a strong laser field QED cascade can be self-sustained and seeded by, for example, an electron at rest. In this case the seeded electron is accelerated in the laser field and emits high-energy photons, which, in turn, create electron-positron pairs as a result of Breit-Wheeler process~\cite{Breit34}. The secondary particles are also involved in photon emission and pair photoproduction, thus, the cascade develops with avalanche-like production of electron-positron pairs and gamma-quanta. The cascade development is very similar to another physical phenomenon, namely, an avalanche ionization in a gas discharge~\cite{raizer1997gas}. Extensive studies of microwave breakdown in gases revealed complex discharge dynamics accompanied by plasma production and generation of gas breakdown waves~\cite{bollen1983high,semenov1982breakdown}. The analogy between a vacuum pair production and a gas ionization, as well as between a vacuum breakdown via QED cascading and a gas breakdown has a deep physical origin~\cite{dunne2012,narozhny2015quantum,efimenko2018extreme}.   

Similar to gas discharges, where the self-sustained and non-self-sustained regimes are possible, there are also two types of QED cascades: the self-sustained QED cascades (or A (Avalanche)-type cascades \cite{mironov2014collapse}), where the laser field provides the energy for cascading, there are also 'shower-like' cascades (or S-type cascades \cite{mironov2014collapse}), where the cascade energy does not exceed the energy of the seed particles. The S-type cascades are well-known as an air showers produced by cosmic rays in atmosphere \cite{rao1998extensive}. The QED cascades of the both types play an important role in astrophysical phenomena ~\cite{rao1998extensive,nerush2017weibel,timokhin2010time}. 

A number of papers are devoted to the search of optimal laser field configurations in order to facilitate experimental realization of QED cascading. Most of the proposed configurations are based on multiple colliding laser pulses~\cite{Bell2008,bulanov2010multiple,gonoskov2012dipole,gonoskov2013probing,bashmakov2014effect,muraviev2015generation,gelfer2015optimized,grismayer2017seeded} or on tightly focused laser pulse~\cite{mironov2017observable} which field structures are strongly different from a traveling plane wave. The plane wave-like configurations are considered as not suitable for cascading because an electron initially at rest cannot be accelerated in this field in such way that it will be capable of emission of gamma-quanta with high enough energy~\cite{rmp,narozhny2015quantum,mironov2017observable,Bulanov13a}. 

\begin{figure}
\includegraphics[width=175mm]{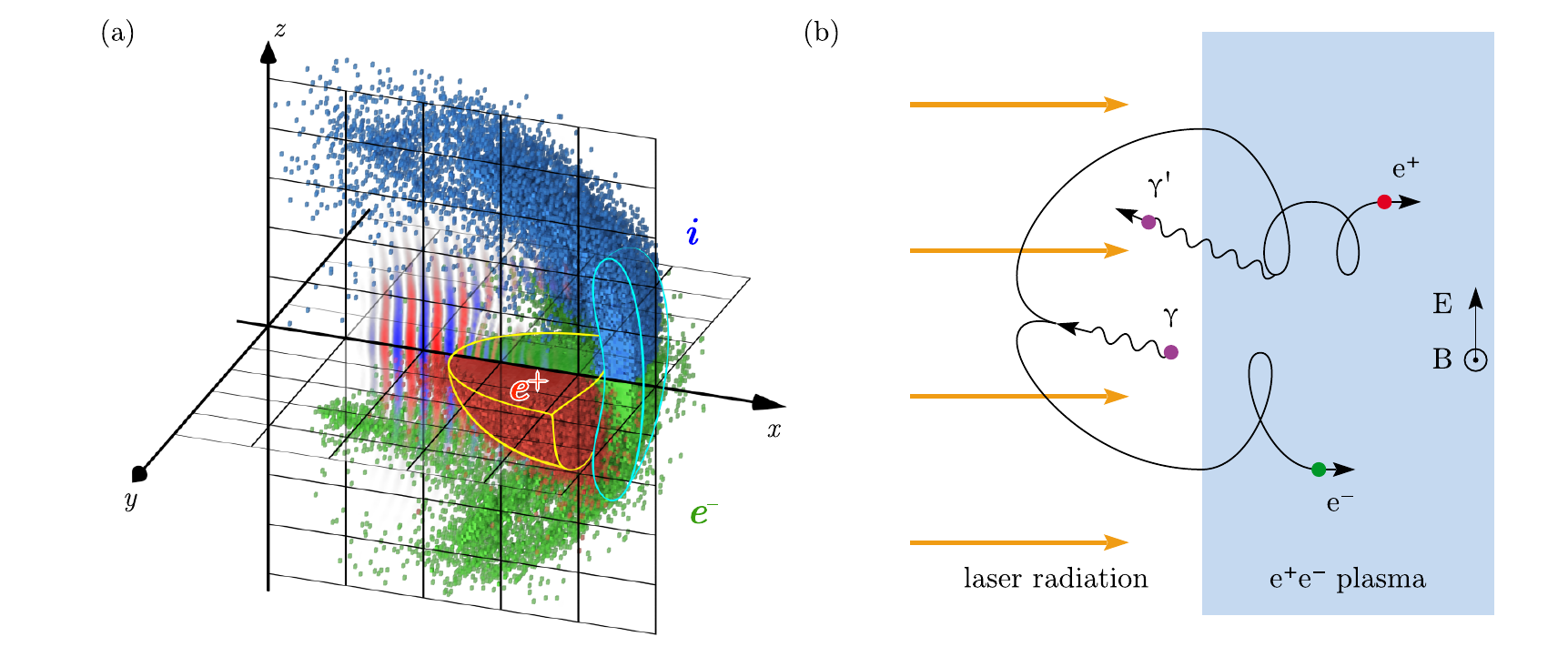} 
\caption{\label{Scheme} (a) The particle distributions (blue~---~ions, green~---~electrons, red~---~positrons) calculated in the simulation at $t=22\,\lambda/c$. The cyan contour roughly outlines the initial foil particles, yellow~---~particles of the electron-positron plasma. The distribution of the $y$-component of the electric field is plotted in $xz$-plane, where red color indicates positive value, blue~---~negative. (b) The schematic of the particle multiplication in the QED cascade in a plane wave. In the vacuum region $E = B$, and $B > E$ in the plasma. The emitted photon $\gamma$ in the laser field produces the electron-positron pair which is pushed to the plasma by the radiation pressure. The electron and the positron in the plasma region can emit the gamma-quanta, e.g. the positron emits the photon $\gamma'$ moving towards the laser field and the proccess repeats.}
\end{figure}

\begin{figure*}
	\includegraphics[width=1\linewidth]{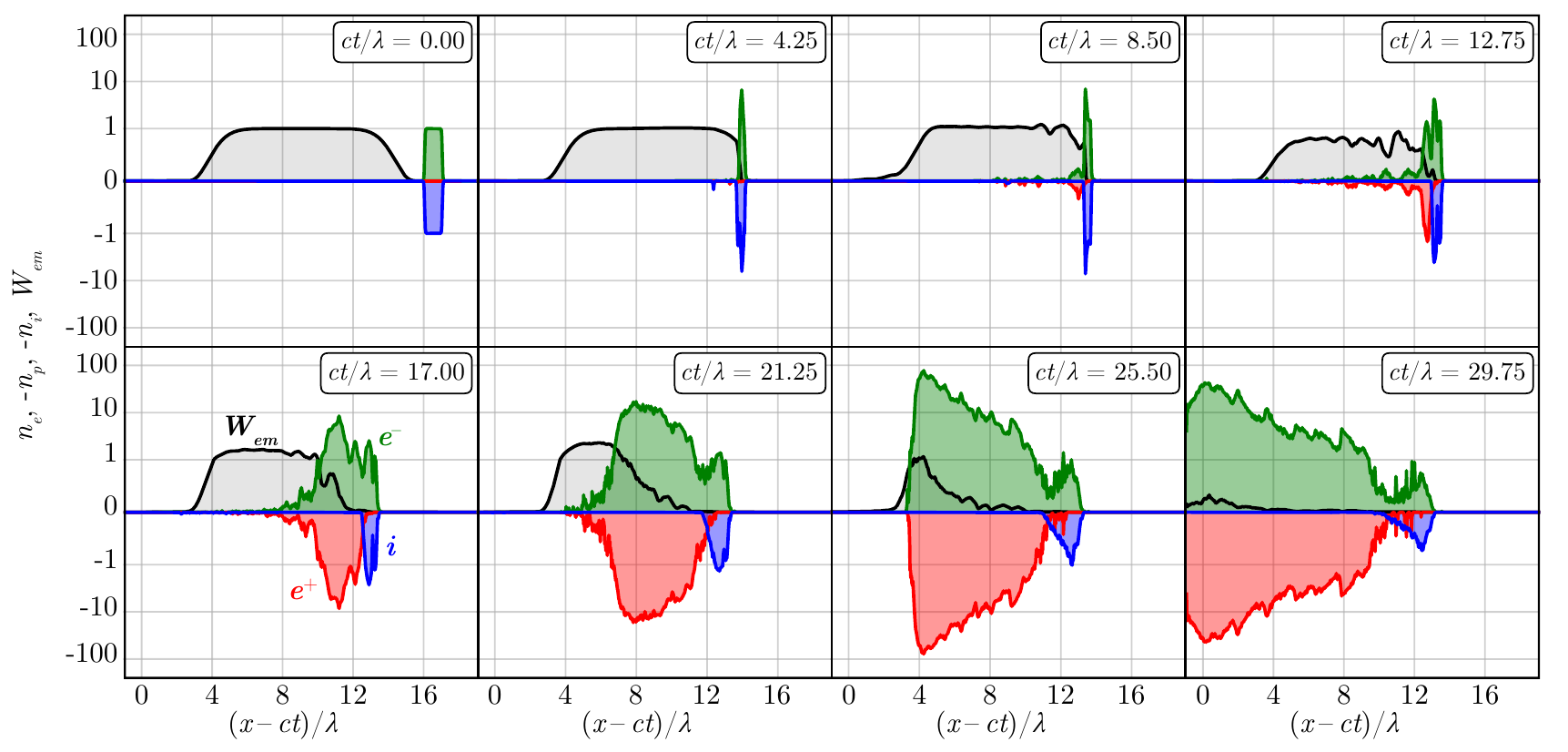} \caption{\label{Sim_res} The distribution of the particle densities (electrons~---~green line, ions~---~blue line, positrons~---~red line) and the electromagnetic energy density (black line) along the $x$-axis for different time moments. The values are normalized to its initial maximums, except positrons whose density is normalized to the maximum of the initial electron density. The scale of the vertical axis is linear for the range $[-1,1]$ and logarithmic for the ranges $[-100,-1]$ and $[1,100]$.}
\end{figure*}

The QED cascade can also develop in the field structure formed by laser radiation interacting with solid targets~\cite{Kostyukov2016}.  The impact of QED effects on the laser-solid interaction are mainly explored in the hole boring regime~\cite{Wilks1992,macchi2013ion}, when the target thickness is much greater than the skin depth characterizing the laser penetration into the foil.  Particularly, the production of the electron-positron plasma `cushions'~\cite{kirk2013pair} is observed in numerical simulations~\cite{Ridgers2012,nerush2015laser,Kostyukov2016,del2017ion} which, however, do not stop the intrinsic ion acceleration. The hole boring regime generally is characterized by significant reflection of the incident laser radiation, and thus cascading can be considered as in a kind of a scheme with two counter-propagating laser pulses. This is not the case for the light sail (LS) regime when the target is a thin foil with the skin depth of the order of the target thickness~\cite{esirkepov2004,macchi2013ion}. In this case the foil can be continuously accelerated as a whole and the laser reflection is negligible. Though LS regime has gained substantial interest during the past years as one of the most efficient schemes for high-energy ion acceleration~\cite{PhysRevSTAB.16.011303}, it is not yet explored at extremely high laser intensities when the QED effects play a key role. 

In this paper we report on a new effect, namely on a vacuum breakdown wave propagation via a self-sustained development of a QED cascade in an extremely intense plane electromagnetic wave where it is generally believed that such cascades are suppressed. The field strength of the wave is much below the Sauter--Schwinger threshold for vacuum pair production~\cite{Sauter31, Schwinger51} but slightly higher than the threshold for self-sustained cascade development in counter-propagating waves~\cite{grismayer2017seeded}. The effect is observed in QED-PIC simulations of the extremely high intensity laser-foil interaction in the LS regime. It is demonstrated that the laser-driven vacuum breakdown accompanied by cascade development leads to production of an overdense electron-positron plasma cushion between the laser radiation and the moving foil. The breakdown front propagates towards the laser radiation in the foil reference frame that qualitatively resembles gas breakdown waves propagating towards the microwave source during microwave discharges in gases \cite{bollen1983high,semenov1982breakdown}. The produced electron-positron plasma absorbs the laser radiation and decouples the radiation from the foil plasma thereby interrupting the ion acceleration. The suppression of LS ion acceleration in the extremely intense laser radiation is another important result of the paper. The cascade continues to develop even after the laser field is decoupled from the foil plasma, thus, the latter can be considered as a seed which becomes negligible for the late stage of the cascade development. 

\section*{Results}
\begin{figure}
\includegraphics[width=175mm]{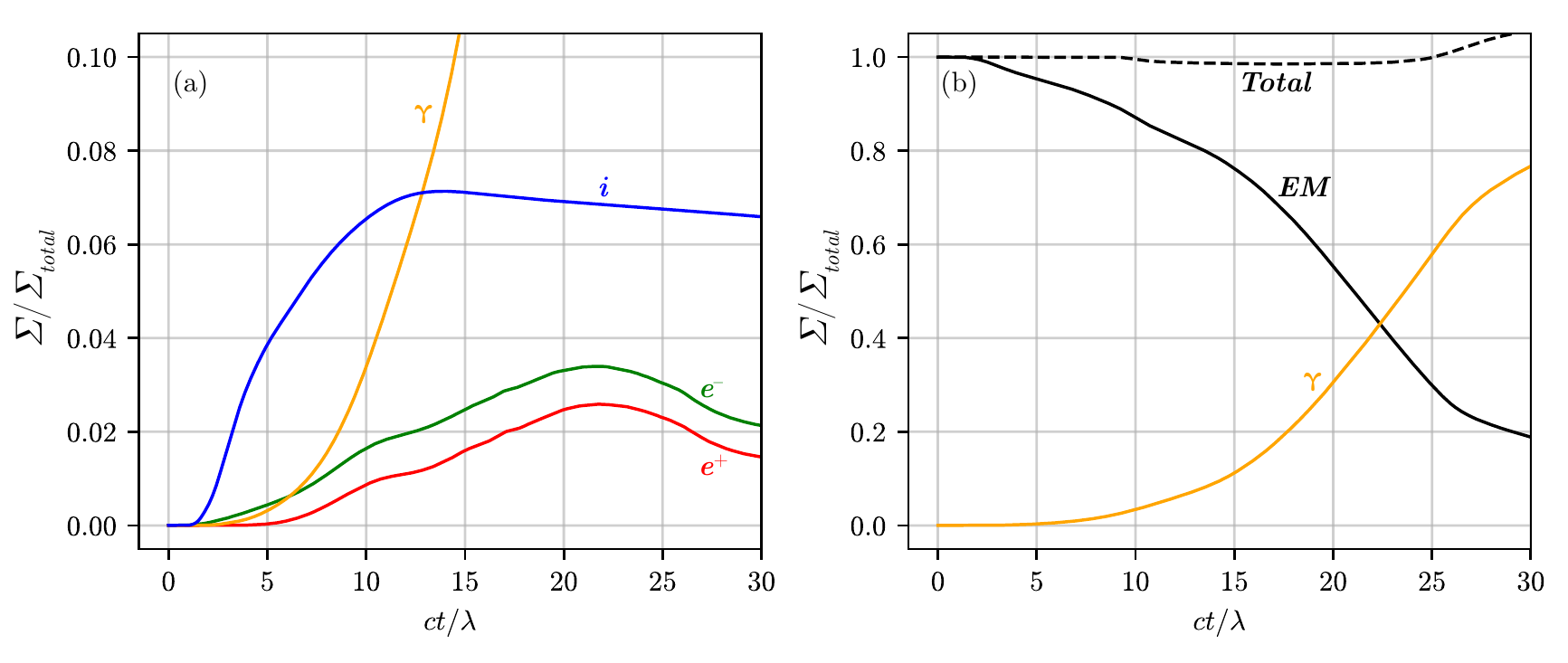} 
\caption{\label{Sim_en} (a) The portion of the laser energy converted into the energy of the gamma-quanta, ions ($i$), electrons ($e^-$) and positrons ($e^+$), respectively as functions of time. (b) The total energy, the ratio of the laser energy to the initial laser energy ($EM$), the portion of the laser energy converted into the energy of the gamma-quanta ($\gamma$), respectively as functions of time. The parameters of the simulation are $n_{e}=5.9\cdot10^{23}~\text{cm}^{-3}$, $d=1~\mu\text{m}$ and $a_{0}=2500$.}
\end{figure}

\begin{figure}
\center{\includegraphics[width=165mm]{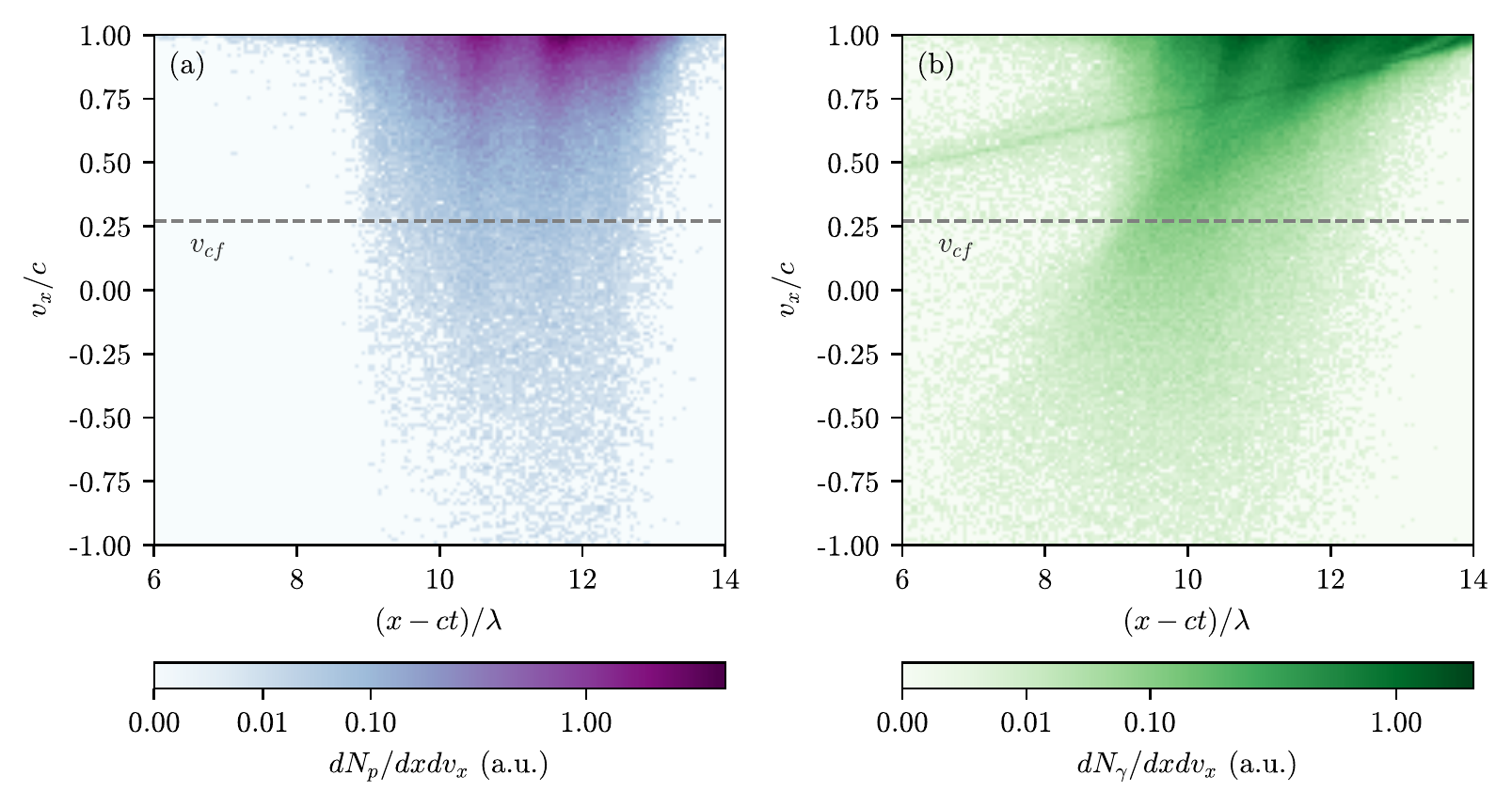}}
\caption{\label{gamma_dist} The distribution of particles [(a)~---~positrons, (b)~---~gamma-photons] in the $x - v_{x}$ space at $t = 20 \lambda/c$. The cushion front velocity is shown with dashed line.}
\end{figure}

\begin{figure*}
	\includegraphics[width=1\linewidth]{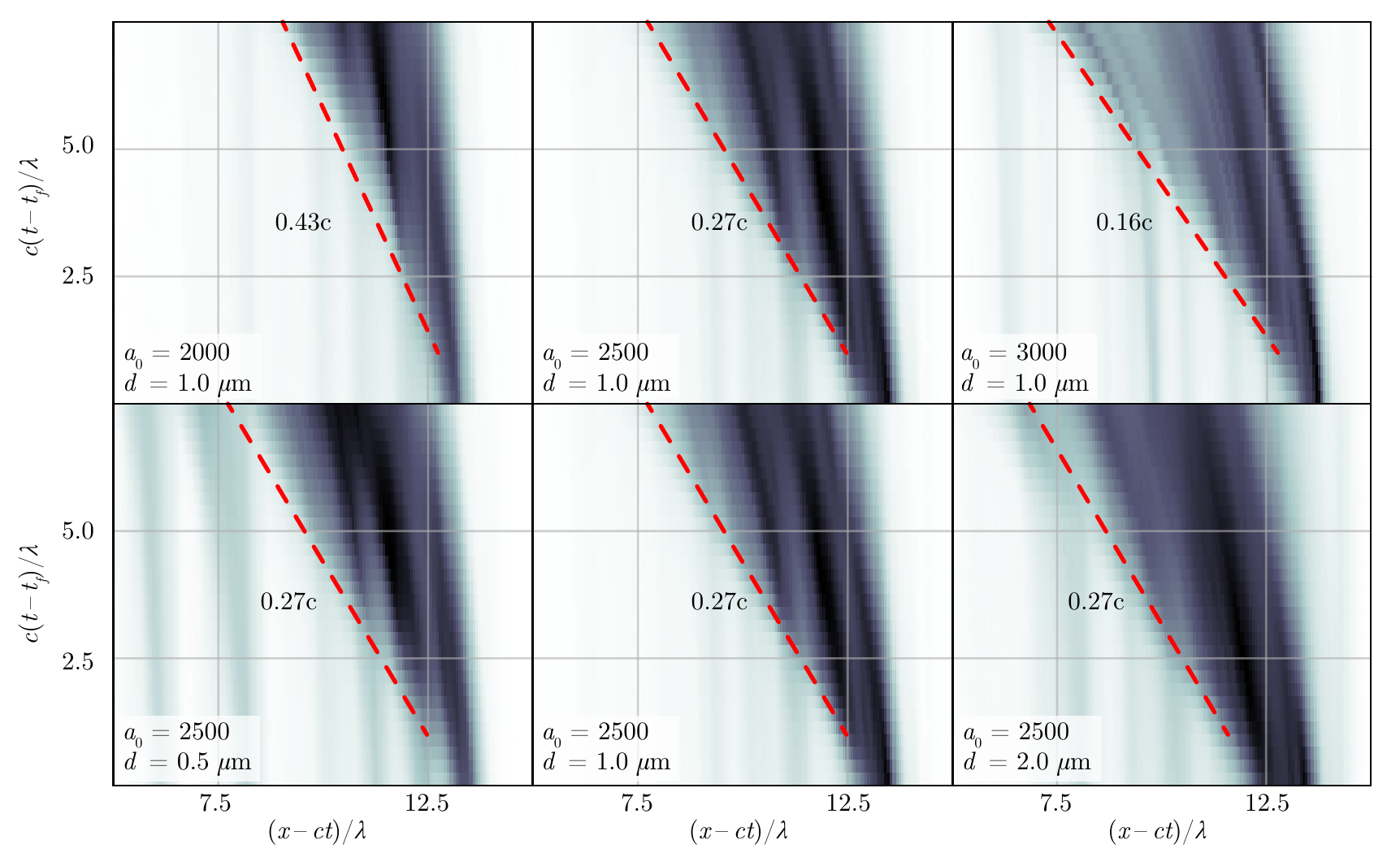}
    \caption{\label{Sim_X(t)} The distribution of the positrons in the plane $x-t$. Darker colors indicate higher particles density. The values of cushion front velocity (red dashed lines) are given in the laboratory reference frame. $t_{f}$ is the time at which the cushion starts to build up. $ct_{f}/\lambda\approx17.5$ for $a_{0}=2000$, $ct_{f}/\lambda\approx15.0$ for $a_{0}=2500$, $ct_{f}/\lambda\approx12.5$ for $a_{0}=3000$.}
\end{figure*}

The typical laser-foil interaction structure observed in the simulations (see Methods for details of the simulation setup) is shown in Fig.~\ref{Scheme}\;(a). The time evolution of the particle density is demonstrated in Fig.~\ref{Sim_res} (see also Supplementary video 1~\cite{Suppl1}) for the simulation with parameters: $n_{e}=5.9\cdot10^{23}~\text{cm}^{-3}\approx530 n_{cr}$ ($n_{cr}=m_{e}\omega_{L}^{2}/4\pi e^{2}\approx10^{21}\text{ cm}^{-3}$ is the critical density), $d=1~\mu\text{m}$, $a_{0}=2500$, where $n_e$ is the initial target electron density, $d$ is its thickness and $a_0$ is the initial maximum of the electric field amplitude. Transverse spatial size of the laser pulse is $10.4 \times 10.4\ \mu$m and its duration is 45 fs. Transverse spatial size of the target is $10 \times 10\ \mu$m. It follows from the simulations that at $ct/\lambda\lesssim 8$ the foil plasma is compressed into a thin layer (see Fig.~\ref{Sim_res}) reflecting the laser radiation while the ions are continuously accelerated, as seen from Fig.~\ref{Sim_en}\;(a). At  this stage when the cushion is absent, and the velocity of the electron-ion foil is much less than the speed of light there is a strong reflection of the incident laser radiation from the foil plasma. As a result, QED cascade efficiently develops in the field of the incident and the reflected plasma waves (see Ref.~\cite{Kostyukov2016}). 

In the time interval $8\lesssim ct/\lambda\lesssim 14$ the non-uniform electron-positron plasma cushion starts to build up. 
As we show later, the magnetic field is stronger inside the electron-positron plasma cushion than the electric one, that leads to complex motion of the electrons and positrons including high-frequency cyclotron rotation (see the positron tracks in the Supplementary video~2~\cite{Suppl3}). As a result, the particles have a very broad distribution of longitudinal velocity [see positron velocity distribution in Fig.~\ref{gamma_dist}\;(a)] and thus some of them are capable to emit gamma-quanta which are overtaken by the cushion front or even counter-propagate the laser pulse [see Fig.~\ref{gamma_dist}\;(b)].
These gamma-quanta decay in the laser field and create electron-positron pairs in front of the foil plasma in the vacuum region. The produced pairs are pushed by laser radiation to the foil plasma thereby forming the electron-positron cushion [see the schematic of the cushion formation mechanism in Fig.~\ref{Scheme}\;(b)]. At this stage the foil velocity is of the order of the speed of light, the laser absorption in the cushion dominates while the laser reflection becomes negligible.

When the cushion becomes dense and thick the laser field is almost completely screened at the location of the foil particles, thus ion acceleration is suppressed [see Fig.~\ref{Sim_en}\;(a)]. In the time period $14\lesssim ct/\lambda\lesssim 28$ the QED cascade develops in the self-sustained regime despite the fact that the laser radiation is decoupled from the foil plasma and no reflection of the laser field is present. As the cascade front propagates with velocity $v_{cf}<c$, the plasma cushion expands towards the laser until all laser energy is absorbed by the produced electron-positron plasma. We call this latter stage the vacuum breakdown wave propagation. It follows from the simulations that up to 70\% of the laser energy is absorbed at $ct/\lambda = 25$ while about 5\%, 7\% and 58\% of the laser energy are deposited into the pairs, the ions and gamma-quanta, respectively [see Fig.~\ref{Sim_en}\;(b)]. Small portion of the particles leave the simulation box that leads to the slight reduction in the pair energy and the ion energy, respectively, at $ct/\lambda\gtrsim 23$.

We perform simulations for different values of $a_{0}=1500$, $2000$, $2500$, $3000$ for $d=1~\mu\text{m}$ and for different values of $d=0.5~\mu\text{m}$, $1~\mu\text{m}$, $2~\mu\text{m}$ for $a_{0}=2500$. It follows from our simulations that the laser energy portions converted into the pair energy and gamma-quanta energy at $ct_{f}/\lambda = 25$ are $\eta_{pair} \equiv \Sigma_{pair}/\Sigma_{total} \simeq 5.1\%$ and $\eta_\gamma \simeq 24.7\%$, respectively, for $a_0 = 1500$; $\eta_{pair}  \simeq 7.2\%$ and $\eta_\gamma \simeq 50.6\%$, respectively, for $a_0 = 2000$; $\eta_{pair} \simeq 6.0\%$ and $\eta_\gamma \simeq 76.7\%$, respectively, for $a_0 = 2500$.  For $a_0 = 3000$ we obtain $\eta_{pair} \simeq 5.7\%$ and $\eta_\gamma \simeq 51.8\%$ at $ct_{f}/\lambda = 20$. It is seen from Fig.~\ref{Sim_X(t)} that the cascade front velocity depends little both on time and on the foil thickness while it strongly depends on the laser intensity and decreases with the increase of $a_0$. It is interesting to note that the electron-positron plasma density at the late stage of the interaction is several times higher than the relativistic critical density~$a_0 n_{cr}$. In all the simulations, the cascade develops efficiently, however, for $a_0 = 1500$ the positron density reached value of about $0.6 a_0 n_{cr}$ at the end of the simulation ($t = 30 \lambda/c$). Thus we suppose that $a_0 = 1500$ is somewhat close to the threshold for an avalanche-type cascade in a plane wave.

In order to understand the role of the foil ions in the QED cascade we simulate an interaction of the laser pulse with an overdense electron-positron plasma layer initially at rest (without any ions). The laser pulse is the same as in the simulation for the foil. The layer parameters: $d=1~\mu\text{m}$ and $n_e = 0.7 a_0 n_{cr}$. It is observed in the simulation (see Supplementary video 3~\cite{Suppl2}) that the self-sustained cascade develops in a similar way as in the case of the electron-ion plasma foil, and the cushion front velocity is the same. Therefore the self-sustained QED cascade can develop in a plane electromagnetic wave if there is an appropriate seed and the wave is intense enough.

\section*{Discussion}
\subsection*{Breakdown front propagation model}
\label{front_model}

The mechanism of the breakdown front propagation is determined by the electron-positron pair photoproduction from the high-energy gamma-quanta which are overtaken by the cushion front [see Fig.~\ref{Scheme}\;(b)]. These gamma-quanta are emitted by the relativistic electrons and positrons inside the cushion and have the greatest photoproduction probabilities. They can get to the vacuum region and produce new electron-positron pairs in the laser field. The laser field then accelerates the created electrons and positrons towards the cushion, where they can emit gamma-quanta which again will lag behind the front, and so on. Thus, the self-sustained cascade develops on the interface of the vacuum region and the cushion. This leads to constant growing of the electron-positron plasma cushion towards the laser. A simple phenomenological model based on discussed above mechanism can be constructed to describe propagation of the breakdown front in the reference frame moving along the $x$-axis with the average electron-positron plasma velocity $v_{pl}$.

We assume that (i) the electrons and positrons are immobile on average but have enough energy to produce gamma-quanta; (ii) they have the same density, $n_{p}=n_e$; (iii) the gamma-quanta are propagating in exactly opposite to the direction of the $x$-axis and (iv) produce the electron-positron pairs; (v) the laser intensity is constant and uniform. In this case the continuity equations can be written as follows:
\begin{eqnarray}
	\label{dnpdt}
  \frac{\partial n_{p}}{\partial t}  & = &   W_{p}n_{\gamma}, \\
	\label{dngdt}   
   \frac{\partial n_{\gamma}}{\partial t}-c\frac{\partial n_{\gamma}}{\partial x} &  = &  -W_{p}n_{\gamma}+2W_{r}n_{p}
   \label{continuity}
\end{eqnarray}
where $n_{\gamma}$ is the gamma-quanta density, $W_{p}$ and $W_{r}$ are the mean probabilities of the pair photoproduction and of the photon emission, respectively. The probabilities are assumed to be constant according to the assumption (v). If the term with $\partial_x$ characterizing the spatial dispersion is neglected in Eqs.~(\ref{continuity}) then they are reduced to the equations describing QED cascade in the rotating electric field without spatial dynamics~\cite{bashmakov2014effect,grismayer2017seeded}. The reduced equations for the rotating electric field configuration can be derived from the self-consistent kinetic equations under the formulated above assumptions~\cite{kostyukov2018growth}.

Eqs.~(\ref{continuity}) can be solved with the one-sided Fourier transform \cite{pitaevskii2012physical}, i.e. by the expansion of their solution as a series of complex exponents with real $k$ values and complex $\omega$ values
\begin{equation}
n_{p,\gamma}\left(t,x\right) = \intop_{-\infty+i\sigma}^{+\infty+i\sigma}\mathrm{e}^{-i\omega t}\frac{d\omega}{2\pi}\intop_{-\infty}^{+\infty}\mathrm{e}^{ikx}n_{p,\gamma}\left(\omega,k\right)\frac{dk}{2\pi}.\label{t-x}
\end{equation}
where
\begin{equation}
n_{p,\gamma}\left(\omega,k\right) = \intop_{0}^{+\infty}\mathrm{e}^{i\omega t}dt\intop_{-\infty}^{+\infty}\mathrm{e}^{-ikx}n_{p,\gamma}\left(t,x\right)dx,\label{w-k}
\end{equation}
and $\sigma$ is a real number so that the contour path of integration is in the region of convergence. 

\begin{figure*}
	\includegraphics[width=1\linewidth]{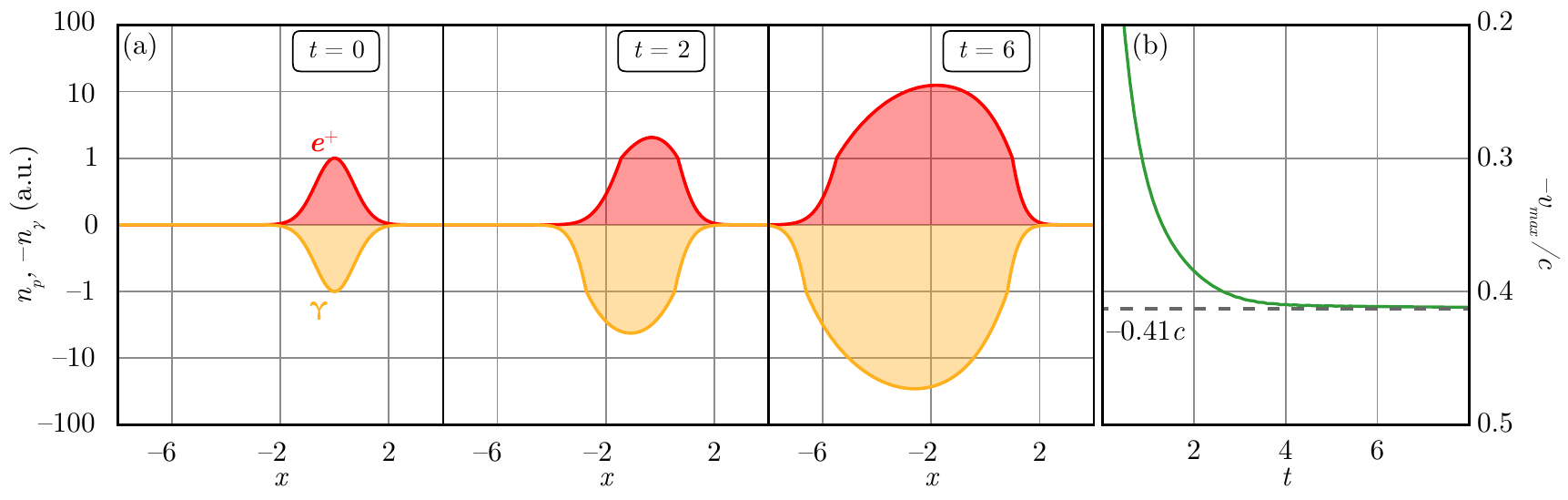}
	\caption{\label{front} Numeric solutions of Eqs. \eqref{dnpdt}~--~\eqref{dngdt}: (a) distribution of the positrons $n_p$ (red line) and the gamma-quanta $n_\gamma$ (orange line) for different time moments. The scale of the vertical axis is linear for the range $[-1,1]$ and logarithmic for the range $[-100,-1],~[1,100]$. Coordinates, time and densities are normalized in such way that $W_r=1.0$, $W_p=0.25$.
(b) Velocity of the maximum of the positron density $n_p$ as a function of time derived from the numerical solution.}
\end{figure*}

If the initial distribution of the pair plasma density and the gamma-quanta density are
$n_{p}\left(0,x\right)$ and $n_{\gamma}\left(0,x\right)$, respectively, then the solution for the pair plasma density is
\begin{equation}
\label{np}
n_{p}\left(t,x\right)  =  \intop_{-\infty+i\sigma}^{+\infty+i\sigma}\frac{d\omega}{2\pi}\intop_{-\infty}^{+\infty}\frac{dk}{2\pi}\intop_{-\infty}^{+\infty}dx' 
  \frac{W_{p}n_{p}(0,x')+i\left(\omega+kc\right)n_{\gamma}(0,x')}{\Delta\left(\omega,k\right)} \mathrm{e}^{ik(x-x')-i\omega t}, 
\end{equation}
where $\Delta\left(\omega,k\right)=\omega^{2}+\omega(kc + iW_{p})+2W_{p}W_{r}$. The initial distribution perturbations propagates along characteristics determined by the dispersion relation $\Delta\left(\omega,k\right)=0$ with the following solution: 
\begin{equation}
\omega=\frac{-kc-i W_p \pm\sqrt{(kc+i W_p)^{2}-8 W_pW_r}}{2}. 
\label{dispersion}
\end{equation}
The group velocity of the perturbations then can be found~\cite{pitaevskii2012physical}, $v_{gr}=\partial\mathrm{Re}[\omega]/\partial k$. In the absence of the laser field the QED processes are suppressed, $W_{r} = W_p=0$. In this case the plasma and the gamma-quanta are decoupled from each other and the dispersion relation yields $\omega=-k$ for the gamma-quanta and $\omega=0$ for the plasma, which correspond to $v_{gr}=-c$ and $v_{gr}=0$, respectively.

It follows from the analysis of the dispersion relation that (i) in the case of large wave numbers $k~\gg~W_{r}/c~>~W_{p}/c$ the perturbations are damped since $\mathrm{Im}\left[\omega\right]=-W_{p}/2<0$; (ii) the perturbations with small wavenumbers, $k\ll W_{p}/c$, are the most unstable with the growth rate 
\begin{equation}
\Gamma \equiv \mathrm{Im}\left[\omega\right] =\frac{W_p}{2} \left(\sqrt{1+\frac{8 W_r}{W_p}}-1\right). 
\label{gamma}
\end{equation}
This growth rate coincides with the QED cascade growth rate derived in Refs.~\cite{bashmakov2014effect,grismayer2017seeded} for the field configuration corresponding to the rotating electric field. From Eq.~\eqref{dispersion}, the dispersion relation for the unstable perturbations is
\begin{eqnarray}
\omega & \approx & \frac{\mu-1}{2}kc + i\Gamma 
\\
\mu & = & \frac{1}{\sqrt{1+8 W_r/W_p}}.
\label{omega}
\end{eqnarray} 
The parameter $\mu $ peaks in the strong QED limit $a_0 \rightarrow \infty$, $W_r/W_p \approx 4$ and $\mu \approx0.17$ \cite{berestetskii1982quantum}. Therefore, in the case of small $k$ the group velocity of the unstable perturbations is $v_{gr}\approx -0.41c$. These results coincide well with the results of numerical solution of Eqs.~(\ref{continuity}) for various shapes of the initial seed of pairs and gamma-quanta (see Fig.~\ref{front}).

In the reference frame moving along with the pair plasma, the cascade front velocity $v_{cf}$ should coincide with the above-mentioned group velocity, $v_{gr} \approx -0.41 c$. In the laboratory reference frame the relation between the cascade front velocity and the mean cushion plasma velocity along the $x$-axis, $v_{pl}$, relates by the Lorentz transform 
\begin{equation}
    v_{cf}=\frac{v_{pl}+v_{gr}}{1+ v_{pl} v_{gr} /c^2} 
\label{vcf}
\end{equation}
$v_{pl}$ can be found from Eq.~(\ref{vcf}):
\begin{equation}
    \label{vpl}
    v_{pl}=\frac{v_{cf}-v_{gr}}{1-v_{gr}v_{cf}/c^2} 
\end{equation}
These reasoning predict $v_{pl}=0.61c$ for $v_{cf} = 0.27 c$ ($a_{0}=2500$, see Fig.~\ref{Sim_X(t)}) that is reasonably close to the value of the averaged positron velocity $\approx 0.75 c$ retrieved from the simulation [see Fig.~\ref{Sim_field}(b)].

\subsection*{Electrodynamics of cascade plasma}
\label{electrodynamics}
\begin{figure}
	\center{\includegraphics[width=85mm]{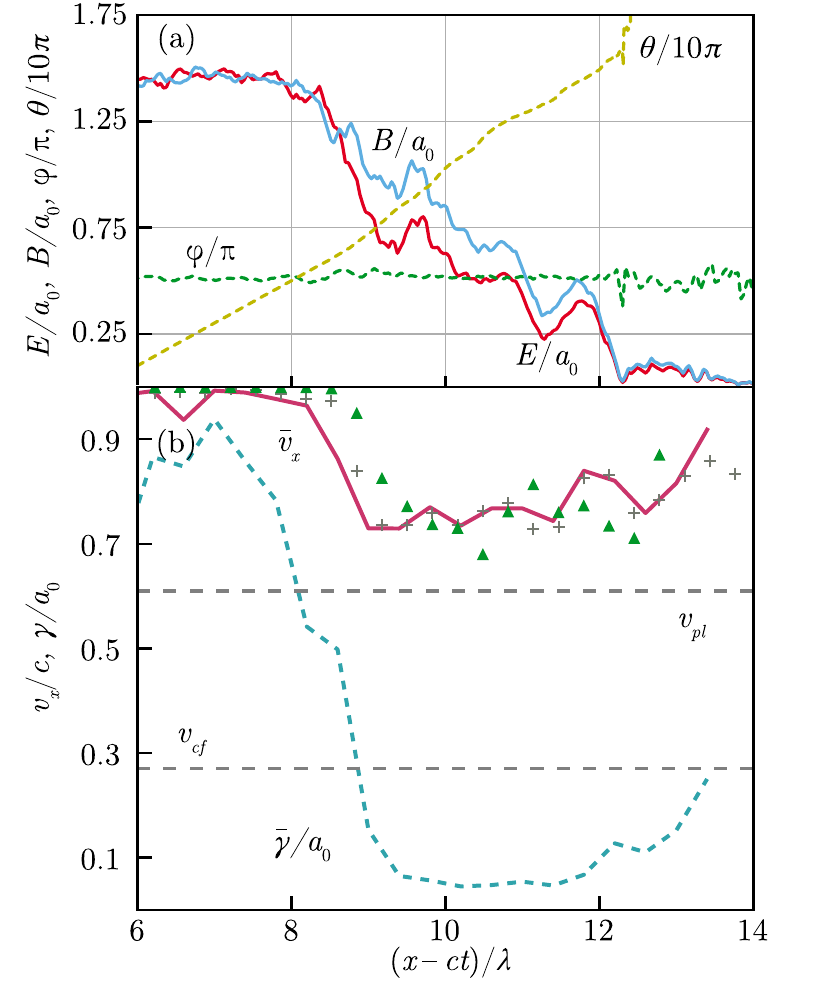}}
    \caption{\label{Sim_field} Results of the numerical simulation at $t = 20 \, \lambda / c$. The simulation parameters are the same as for Fig.~\ref{Sim_en}. The value of electric, $E$, and magnetic, $B$, fields perpendicular to the $x$-axis, and the angle between them, $\varphi$. The angle $\theta$ is between the electric field and the $y$-axis counted counterclockwise in the $yz$-plane. (b) The positrons located in $\sim 1 \, \lambda$ neighbourhood of the laser pulse axis:
their mean velocity along the $x$-axis, $\bar v_x$, (solid line) and their mean Lorentz factor $\bar \gamma$ as functions of $x$. The drift velocity $(\mathbf{E \times B})_x / B^2 $ (plus signs) and the mean positron velocity found from the positron density and the electric field with Eq.~\eqref{vM} (triangles). The cascade front velocity $v_{cf}$ obtained from the simulation and the velocity of the pair plasma $v_{pl}$ predicted by Eq.~\eqref{vpl}.}
\end{figure}

To characterize the cascade in more detail the EM field distribution and the particle dynamics inside the electron-positron plasma in the cushion are studied. It follows from Fig.~\ref{Sim_field}(a) that the field structure is close to a circularly polarized wave with perpendicular electric and magnetic components of the field, $\mathbf{E \perp B}$, and the field declines in the plasma within several laser wavelenghts. The key feature of the EM field is the magnetic field predominance, $B > E$, inside the electron-positron plasma. In such field the electrons and positrons do not gain energy [see the line $\bar \gamma / a_0$ in Fig.~\ref{Sim_field}(b)], thus the self-sustained cascading is suppressed deep inside the cushion. Note that probable signatures of a cascade in a plane wave are also observed in the simulations for the case of linear polarization~\cite{muraviev2015generation}, and are also accompanied by a magnetic field dominance.

We develop simple model based on the Maxwell equations, the asymptotic 	theory~\cite{gonoskov2018radiation, Samsonov18a} and assumptions supported by the numerical simulations. The asymptotic theory states that for high enough field strength the charged particle is attracted to the "asymptotic" trajectory~\cite{Samsonov18a} due to the radiation reaction or, in other words, moves almost along "radiation-free 	direction"~\cite{gonoskov2018radiation} providing minimal radiative losses. The particle velocity along the asymptotic trajectory can be 	calculated from algebraic equations depending on the local field strength.  As compared to standard  equations of motion in the differential form, the algebraic equations strongly simplify 	calculations. 	Generally, there is a small deviation angle between the radiation-free direction and the particle velocity caused by dependency of the radiation-free direction on time and by stochastic nature of radiation 	reaction~\cite{gonoskov2018radiation}. The resulting deviation angle is significant enough to produce copious gamma-quanta.

We start from the Maxwell equations
\begin{eqnarray}
\label{Maxwell1}
\nabla \times \mathbf E &=& -\frac{1}{c} \frac{\partial \mathbf B}{\partial t}, \\
\label{Maxwell2}
\nabla \times \mathbf B &=&  \frac{1}{c} \frac{\partial \mathbf E}{\partial t} + \frac{4 \pi}{c}
\mathbf j,
\end{eqnarray}
where the particle velocity determining the current is taken from the asymptotic theory. The Maxwell equations should be supplemented by the continuity equation for the plasma density and the current density $\mathbf j$. For simplicity we neglect the laser field absorption and consider a high-intensity circularly polarized plane wave travelling along the $x$-axis in a homogeneous electron-positron plasma with constant amplitude, hence the average electron (positron) velocity along the $x$-axis and the plasma density remains constant.
	
Accordingly to Fig.~\ref{Sim_field}\;(a) we suppose that the electric and magnetic fields are almost perpendicular to each other, however the Lorentz invariant $\mathbf{E \cdot B} < 0$ [in Fig.~\ref{Sim_field}\;(a) $\varphi$ is slightly greater than $\pi/2$]. Thus, there is a reference frame $K'$ moving along the $x$-axis with the speed $(\mathbf{E \times B})_x / B^2 \approx E / B$ in which the component of the electric field perpendicular to the magnetic field vanishes, hence the electric field is directed exactly opposite to the magnetic one. 
In accordance with the asymptotic theory~\cite{Samsonov18a} the induced current is parallel to the electric field as well as magnetic one in the reference frame $K'$. Hence, in the laboratory reference frame the current density is $\mathbf j = -2 e n_p \bar v_\perp \mathbf B / B$, where $n_p$ is the positron density (half of the overall plasma density) and $\bar v_\perp$ is the average positron velocity perpendicular to the $x$-axis.

We assume that in the circularly polarized wave the magnetic and electric fields $\mathbf B$ and $\mathbf E$ rotate
counterclockwise in the $yz$-plane with the increase of $x$ [see Fig.~\ref{Sim_field}\;(a)] and clockwise with the increase of $t$. Thus, for the field derivatives we have: 
\begin{eqnarray}
    &\nabla \times \mathbf E = -k \mathbf E, \quad
    &\partial_t \mathbf E = -\omega E \mathbf b, \\
    &\nabla \times \mathbf B = -k \mathbf B, \quad
    &\partial_t \mathbf B = \omega B \mathbf e,
\end{eqnarray}
where $\mathbf b = \mathbf B / B$ and $\mathbf e = \mathbf E / E$, that allows the decomposition of Eqs.~\eqref{Maxwell1}~--~\eqref{Maxwell2} on two perpendicular components: one parallel to $\mathbf e$ and the other parallel to $\mathbf b$, that yields:
\begin{eqnarray}
    \label{Maxwell3}
    &&kE = \frac{\omega}{c} B, \\
    \label{Maxwell4}
    &&kB = \frac{\omega}{c} E + 8 \pi e n_p \frac{\bar v_\perp}{c},
\end{eqnarray}
where $\omega$ is the frequency of the wave, $k$ is its wavenumber, hence
\begin{equation}
    \label{dispersion-maxwell}
    c^2 k^2 = \omega^2 + \frac{\omega_{cr}^2 \bar v_\perp}{c a},
\end{equation}
where $a = e E / m_e c \omega$ is the normalized amplitude of the electric field and $\omega_{cr} = (8 \pi e^2 n_p / m_e)^{1/2}$ is the critical frequency.

In order to find the mean particle velocity along the $x$-axis, $\bar v_x \approx E / B$, we consider PIC simulation in which it is seen that $\bar v_\perp$ relates with $\bar v_x$ as follows:
\begin{equation}
    \bar v_\perp = \nu (c^2 - \bar v_x^2)^{1/2},
\end{equation}
where $\nu < 1$ is a numerical coefficient introduced because of the large spread of the particle velocity distribution in the cushion. The simulations show that independently on the laser amplitude and time $\nu \approx 1/6$. Then $\bar v_x$ can be calculated from Eqs.~\eqref{Maxwell3}~--~\eqref{Maxwell4}:
\begin{equation}
    \label{vM}
    \frac{\bar v_x}{c} = \left(\frac{\sqrt{1 + 4 S^2} - 1}{2 S^2} \right)^{1/2}, \quad
    S = \frac{2 n_p \nu}{n_{cr}a}, 
\end{equation}
where $n_{cr} = m_e \omega^2 / 4\pi e^2$. Note that in this solution $\bar v_x$ do not depend neither on $\omega$ nor on $k$.

It is seen from Fig.~\ref{Sim_field}\;(b) that $\bar v_x$ calculated from Eq.~\eqref{vM} for $n_p$ and $E$ retrieved from the simulations [shown in Fig.~\ref{Sim_field}\;(b) with triangles] coincides well with the drift velocity $E/B$ (crosses) and the mean positron velocity computed for PIC quasiparticles (solid red line). Although the electrodynamic model is simple with only one numerical parameter, $\nu$, deduced from the simulations it is capable to reproduce the complex, non-monotonic dependence calculated in PIC-QED simulations with surprisingly good accuracy. This justifies the approach and the assumptions used in the model. 

\begin{figure*}
\center{\includegraphics[width=85mm]{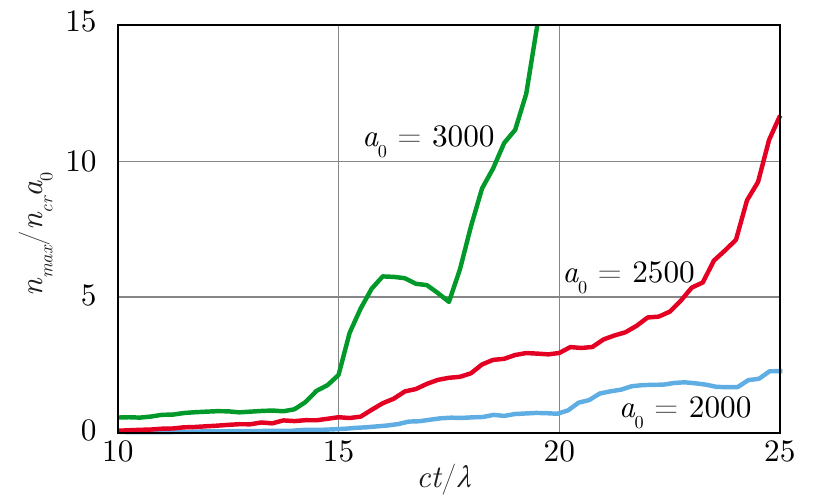}} \caption{\label{Sim_nncr} Maximum value of the parameter $n_p /n_{cr} a_0$ proportional to the parameter $S$ [see Eq.~\eqref{vM}] as a function of time plotted for different values of $a_0$.}
\end{figure*}

Furthermore, in QED plasmas the parameter $S \propto n_p / a_0 $ typically increases with the increase of $a_0$ because of efficient avalanche-like pair production and sharp dependence of the $n_p$ on the $a_0$ (see Fig.~\ref{Sim_nncr}). Thus, according to Eq.~(\ref{vM}), the higher $a_0$, the lower $\bar v_x$. From the front propagation model, independently on the value of $a_0$, the cushion front velocity is always the same in the reference frame moving with the plasma. Hence, as the plasma velocity declines with the increase of $a_0$, in the laboratory reference frame the cushion front velocity also declines with the increase of the laser field strength [see Eq.~\eqref{vcf}], that is in agreement with the simulation results (see the upper row in Fig.~\ref{Sim_X(t)}).

\subsection*{Vacuum breakdown waves}
It is important to note that a plane electromagnetic wave has been generally believed not suitable for self-sustained QED cascading~\cite{rmp,narozhny2015quantum,mironov2017observable,Bulanov13a}.  However, we have shown that cascade can develop in a plane wave in the self-sustained regime if the cascade seed consists of a large enough number of the particles which are capable to properly modify the wave field. It is possible even when the reflection of the incident plane wave by the seed is negligible. The cascading leads to the efficient conversion of the incident low energy photons into the high energy photons and electron-positron pairs, whereas the seed is moving almost with the speed of light. Thus, the cushion of the electron-positron plasma is produced, and the cushion front (the laser-plasma interface) is moving slower than the seed because of continuous production of the electron-positron plasma on the interface (see Fig.~\ref{Sim_res} and Supplemental Material~\cite{Suppl1}). Similarly to ionization waves in the gas discharge physics~\cite{bollen1983high,semenov1982breakdown}, the interface propagation can be considered as a vacuum breakdown shock wave. The cushion plasma density eventually exceeds the relativistic critical density and the plasma screens the seed particles from the incident wave.

The cascade growth rate in the field of two counter-propagating circularly polarized EM waves was calculated numerically in Ref~\cite{grismayer2017seeded} as a function of $a_0$. The threshold value of $a_0$ can be estimated form the condition that the particle number is doubled during the laser period. For 1\;$\mu $m wavelength the threshold value of $a_0$ is about $10^3$. It follows from our simulations that QED cascading accompanied by cushion formation starts when $a_0 > 1500$. Therefore, the vacuum breakdown wave occurs if the field strength of the wave is slightly higher than the threshold for counter-propagating waves, that is much below the Sauter--Schwinger threshold for vacuum pair production $a_0 \simeq m c^2/ \hbar \omega _L \simeq 4 \times 10^5 $~\cite{Sauter31, Schwinger51}. It follows from our simulations that the threshold intensity of incident wave is about $6 \times 10^{24}$~W/cm$^2$ for $1$~$\mu$m wavelength. Interesting features of such cascading is that the cascade front velocity, or in other words, the velocity of the wave-plasma interface can be significantly less than the speed of light and almost does not depend on time. However the front velocity and the mean cushion particle velocity decrease with increasing of the incident wave amplitude. The obtained results are similar for the seed in the form of electron-ion plasma layer and in the form of the electron-positron plasma layer. 

We have developed simple analytical model based on the asymptotic theory of the electron motion in the strongly-radiation-dominated regime~\cite{gonoskov2018radiation, Samsonov18a} that gives insight into the electrodynamics of the cushion plasma.  Another phenomenological model proposed in the paper describes the vacuum breakdown wave propagation and predicts the difference between the cascade front velocity and the mean velocity of the cushion particles that is in agreement with numerical results. Also, the models explain the observed dependence of the cascade front velocity on the laser intensity. In order to provide higher model accuracy additional effects should be taken into account: the dynamics and distribution of the electromagnetic field inside the cushion plasma, the dependence of the probability rates of QED processes on the filed strength and the particle momentum,  the energy distribution of the electrons, positrons and gamma-quanta, etc. 

\section*{Conclusions}

We have demonstrated by 3D QED-PIC simulations of the laser-foil interaction at extremely high laser intensity that (i) a laser-driven vacuum breakdown in a form of QED cascade development is an immanent process for most of high-field phenomena and can develop even in a plane electromagnetic wave, or in other words, in wider range of the field configurations than it was previously supposed; (ii) the LS regime of the ion acceleration becomes inefficient at extremely high intensities when the overdense electron-positron plasma cushion is produced between the laser radiation and the the foil. 

The threshold intensity for the vacuum breakdown can be reached with the upcoming laser facilities, hence the findings of the paper can be important for their applications. The occurrence of the vacuum breakdown wave and consequent light absorption broadens the limitations of the attainable laser intensity~\cite{Bell2008, fedotov2010limitations} to the case of plane-wave geometry. We have demonstrated that the laser-driven ion acceleration scenario is changed dramatically at extremely high laser intensities. Namely, the ion acceleration becomes inefficient because of formation of the cushion decoupling the laser radiation from the ions. However the obtained results can be also used to improve acceleration by, for example, choosing of the appropriate schemes of laser-target interaction which provide suppression of the cushion formation. The developed models can be also applied to the astrophysical phenomena like pair cascade in magnetospheres of neutron stars where the cascading has complex space-time dynamics and can be also accompanied by generation of the vacuum breakdown waves~\cite{timokhin2010time}.

\section*{Methods}

The laser-foil interaction is simulated with 3D QED-PIC code QUILL~\cite{QUILL, QUILL_MC}, which enables modelling of QED effects via Monte-Carlo method. In the simulations the circularly polarized laser pulse with wavelength $\lambda=2\pi c/\omega_{L}=1 \text{ }\mu \text{m}$ propagates along the $x$-axis and has a rectangular profile with smoothed edges along all axes 
\begin{equation}
a(x,y,z,t=0) =  a_0 \cos^2 \left( \frac{ \pi }{2}   \frac{x^4}{\sigma_x^4 } \right) \cos^2 \left( \frac{ \pi}{2}   \frac{y^4}{\sigma_y ^4 } \right) \cos^2 \left( \frac{ \pi}{2}   \frac{z^4}{\sigma_z ^4 } \right)\cos \left( \frac {x \omega_L } {c} \right),
\end{equation}  
where $a_{0}=eE/m_{e}c\omega_{L}$, $\omega_{L}$ is the laser frequency, $m_{e}$ and $e>0$ are the electron mass and charge value, respectively, $c$ is the speed of light. The transverse spatial size of the laser pulse is  $2\sigma_y = 2\sigma_z = 10.4~\mu$m  and the pulse duration is $45$~fs ($2 \sigma_x = 13.4~\mu$m). The laser field structure is very close to a traveling plane wave. The foil plasma of the thickness $d$ and with the initial electron particle density $n_{e}$ has the same transverse size as the laser pulse. The simulation box size is $20\lambda\times30\lambda\times30\lambda$, the grid size is $2000\times300\times300$. Although there is initially one quasi-particle per cell this number significantly rises during the simulations because of the particle production (up to 40 on average and up to  100 at most). We conducted a series of simulations varying parameters $a_{0}$, $n_{e}$ and $d$, but satisfying condition $a_{0}=f n_{e}d\lambda r_{e}$ corresponding to the LS regime \cite{macchi2013ion}, where $r_{e}=e^{2}/m_{e}c^{2}$ is the classical electron radius, $f$ is the numeric coefficient of the order of unity (in all the simulations $f=1.5$). 
It follows from our simulations that the QED cascade dynamics depends on $a_0$ and almost does not depend on the focal spot size if the spot size is much greater than the laser wavelength.

\bibliography{main-10.bbl}

\section*{Acknowledgements}
The work is supported in parts by the Russian Science Foundation by Grant No.~18-11-00210 (numerical simulations), by the RAS Presidium Program  '' Extreme light fields and their interaction with matter '' (model of QED cascade dynamics) and by the Grants Council under the President of the Russian Federation through Grant No. MK-2218.2017.2 (model of pair cushion electrodynamics).

\section*{Author Contributions}
A.S. carried out the simulations and analyzed the results. I. K. developed the front propagation model and wrote the bulk of the manuscript. E. N. developed the model of electrodynamics of the cascade plasma. All authors discussed the results, reviewed and commented on the manuscript.

\section*{Additional Information}
\textbf{Competing interest:} The authors declare no competing interests.

\end{document}